\documentclass[fleqn,usenatbib]{mnras}
\hypersetup{draft}
\usepackage{newtxtext,newtxmath}

\usepackage[T1]{fontenc}
\usepackage{ae,aecompl}

\usepackage{graphicx}	
\usepackage{amsmath}	
\usepackage{amssymb}	

\usepackage{multirow}
\usepackage{multicol}
\usepackage{tablefootnote}

\title[A new white dwarf companion to the $\Delta\mu$ star GJ~3346]{A new white dwarf companion around the $\Delta\mu$ star GJ~3346}

\author[M. Bonavita et al.]{Bonavita, M.$^{1,2}$\thanks{E-mail: mbonav@roe.ac.uk}, C. Fontanive$^{3}$, S. Desidera$^{4}$, V. D'Orazi$^{4}$, A. Zurlo$^{5,6}$, K. Mu{\v z}i{\'c}$^{7}$, \newauthor B. Biller$^{1,2}$, R. Gratton$^{4}$, D. Mesa$^{4}$,  A. Sozzetti$^{8}$ 
\\
$^{1}$ SUPA, Institute for Astronomy, University of Edinburgh, Blackford Hill, Edinburgh EH9 3HJ, UK\\
$^{2}$ Centre for Exoplanet Science, University of Edinburgh, Edinburgh EH9 3HJ, UK\\
$^{3}$ Center for Space and Habitability, University of Bern, Gesellschaftsstrasse 6, 3012 Bern, Switzerland\\
$^{4}$ INAF Osservatorio Astronomico di Padova, Vicolo dell'Osservatorio 5,  35121 Padova, ITALY\\
$^{5}$ N\'ucleo de Astronom\'ia, Facultad de Ingenier\'ia y Ciencias, Universidad Diego Portales, Av. Ejercito 441, Santiago, Chile\\
$^{6}$ Escuela de Ingenier\'ia Industrial, Facultad de Ingenier\'ia y Ciencias, Universidad Diego Portales, Av. Ejercito 441, Santiago, Chile \\
$^{7}$ CENTRA, Faculdade de Ci\^{e}ncias, Universidade de Lisboa, Ed. C8, Campo Grande, P-1749-016 Lisboa, Portugal\\
$^{8}$ INAF - Osservatorio Astrofisico di Torino, Via Osservatorio 20, 10025 Pino Torinese, Italy
}

\date{Accepted 2020 February 21. Received 2020 February 21; in original form 2020 January 27}

\pubyear{2020}

\begin{document}
\label{firstpage}
\pagerange{\pageref{firstpage}--\pageref{lastpage}}
\maketitle

\begin{abstract}
We present the discovery of a white dwarf companion {at $\sim3.6^{\prime\prime}$ from} GJ 3346, a nearby ($\pi\sim42$~mas) K star observed with SPHERE@VLT as part of an open time survey for faint companions to objects with significant proper motion discrepancies ($\Delta\mu$) between Gaia DR1 and Tycho-2. 
Syrius-like systems like GJ 3346~AB, which include a main sequence star and a white dwarf, can be difficult to detect because of the intrinsic faintness of the latter. They have, however, been found to be common contaminants for direct imaging searches. White dwarfs have in fact similar brightness to sub-stellar companions in the infrared, while being much brighter in the visible bands like those used by Gaia.
Combining our observations with Gaia DR2 and with several additional archival data sets, we were able to fully constrain the physical properties of GJ~3346~B, such as its effective temperature (11$\times10^3\pm$500~K) as well as the cooling age of the system (648$\pm$58~Myrs). This allowed us to better understand the system history and to partially explains the discrepancies previously noted in the age indicators for this objects. 
Although further investigation is still needed, it seems that GJ 3346, which was previously classified as young, is in fact most likely to be older than 4 Gyrs.
Finally, given that the mass (0.58 $\pm$ 0.01 $M_{\odot}$) and separation (85 au) of GJ~3346~B are compatible with the observed $\Delta\mu$, this discovery represents a further confirmation of the potential of this kind of dynamical signatures as selection methods for direct imaging surveys targeting faint, sub-stellar companions. 
\end{abstract}



\section{Introduction}\label{intro}

Long-term proper motion measurements provided by historical catalogues like Tycho-2 \citep{tycho} can be a good approximation of the motion of the center of mass of binaries with sufficiently long periods.
Short-term measurements such as the ones provided by the Hipparcos \citep{hip} or, more recently, by the European Space Agency (ESA) cornerstone mission Gaia \citep{Gaia_Main}, can instead capture the reflex orbital motion of the pair. 
A significant difference ($\Delta\mu$) between proper motion measurements can therefore be interpreted as a good indication of the presence of a perturbing body around a seemingly single star.
Targeted searches for companions compatible with measured trends between the Hipparcos and Tycho-2 catalogues \citep[see e.g.][]{makarov2005} have been highly successful \citep[see e.g.][]{tokovinin2012, tokovinin2013}, confirming the power of such selection method. 

The discovery space of these searches is of course limited by the precision of the available measurements, which explains why previous surveys were only able to target stellar-like companions.
The first two intermediate Gaia data releases  \citep[Gaia DR1 and GR2]{GDR1, GDR2} already allow the community to access the processed and calibrated data collected by the spacecraft in its first 22 months of operation. 
The five-parameter astrometric solution based on Gaia data only are now available for for more than 1.3 billion sources, including proper motion measurements with uncertainties below 0.06 mas/yr for the brightest sources. 
As recently demonstrated by \citet{fontanive2019b} with the new COPAINS (Code for Orbital Parametrisation of Astrometrically Inferred New Systems) tool and previously by \cite{Brandt2018}, such an exquisite precision allows to unveil much smaller trends, effectively extending the $\Delta\mu$ searches below the substellar mass limit.

Astrometric signatures could therefore represent a powerful tool to maximise the number of direct detections of wide substellar companions. 
As these objects seem to be rare (see e.g., \citealp{vigan2017}), a carefully pre-selected sample may in fact lead to a higher number of detections, compared to a blind search.
In addition to precise astrometry, Gaia DR2 also provides multi-band photometry for a considerable amount of sources, allowing to better characterise the faint companions detected via Direct Imaging (DI). Gaia colours hence make it possible to identify contaminants such as the so-called Sirius-like systems, composed by a main sequence star of spectral type earlier than M and a white dwarf (hereafter WD) companion \citep{Holberg2013}.
The faint WD would appear very similar to a young planetary or brown dwarf companion in the infrared, while being much brighter in the visible bands surveyed by Gaia. 

Because of the intrinsic faintness and small projected separation of the companions,  the current census of Sirius-like systems is highly incomplete, even within a short distance from the Sun (few tens of pc). 
The use of state-of-the-art high-contrast instrumentation has lead to a number of discoveries in the last few years \citep{zurlo2013, crepp2013, crepp2018}, contributing towards reducing such incompleteness, and confirming the existence of an unseen population of WD companions in the close vicinity of the Sun, as predicted by \cite{Holberg2013}. 

The newly detected companions are typically found at rather small angular separations, corresponding to physical separations of few tens of au, and therefore close enough to have harboured some  accretion  phenomena. This makes them very useful benchmark objects to constrain wind accretion occurring in moderately wide binaries during the AGB phase of the WD progenitor. They can also be used to investigate the maximum binary separation at which Ba-stars can be observed, as well as to characterise the rate of companion loss as a function of orbital periods.

Long-term radial velocity trends are observed in some cases, providing crucial clues towards the determination of dynamical masses. 
This, together with the availability of precise parallax measurements of the host star allows for a calibration of the progenitor mass \citep[see e.g.][]{weidemann2000} and the empirical mass radius relations \citep[see e.g.][]{tremblay2017}, as well as the luminosity function of WDs \citep[see e.g.][]{holberg2016}, although cases of discrepancy between the WD cooling age and properties of the companion have been previously reported \citep[see e.g.][]{Matthews2014}.

Furthermore, some of the central stars show signatures of the impact of mass-loss from progenitors of WDs on the central stars.
These include alterations of the rotation and therefore of the magnetic activity level \citep[][, D'Orazi et al. in preparation]{desidera2007,zurlo2013}, and alterations of chemical abundances of selected elements \citep{jeffries1996b,desidera2016}, which could lead to a mis-classification of the host as young star.
These alterations may occur through direct mass exchange between the components
\citep[Roche lobe overflow, see e.g.][]{mccrea1964,iben1993} or, for wider binaries, by accretion of material lost during the AGB phase through stellar wind \citep[wind accretion, see e.g.][]{jeffries1996,boffin2015}.
This latter mechanism appears to be relatively efficient in providing alterations even for binary separations of several tens of au.
The occurrence of WD with moderately wide companions with chemical alterations 
of s-process elements and carbon \citep{jeffries1996b} shows
unambiguously that the origin of the accreted material is an AGB star (progenitor of the WD).
Accretion of small amounts of material in these cases is also predicted by binary evolution codes \citep{hurley2002}.
Furthermore, it should be considered the original separation of the binary system at the end of the AGB phase would have been smaller than the present one, because
of the significant mass loss experienced by the system \citep{hadjidemetriou1963}.

In this paper, we present the discovery of a WD companion to GJ 3346, one of the targets of the COPAINS pilot survey (Bonavita et. al in prep.).
The observational setup and data reduction is described in Section~\ref{sec:obs}. Section~\ref{sec:host} presents the properties of the host star, and Section~\ref{sec:res} presents the analysis of the properties of the new companion. Finally our results are discussed and summarised in Section~\ref{s:conclusion}.

\section{Observations and data reduction}\label{sec:obs}
GJ 3346 was observed as part of the COPAINS pilot survey (Bonavita et. al in prep.), an open time SPHERE program (ID 100.C-0646) aimed at validating the COPAINS target selection process for direct imaging systems, presented in \citet{fontanive2019b}.
The goal of this study is to image unseen companions to stars selected with the COPAINS tool for their significant proper motion differences ($\Delta\mu$) between the astrometric values from the first Gaia data release and historical proper motions from the Tycho-2 catalogue. We note that the Gaia DR2 catalogue was not yer available at the time the survey was devised. 
Based on predictions from the COPAINS code \citep{fontanive2019b}, the hidden companions responsible for the observed trends in the survey targets were expected to possibly be of sub-stellar and in some cases of planetary nature (see Section \ref{sec:astrometry}). 

In order to enhance our capability to detect such objects, despite their possible low luminosity, we therefore choose to use the SPHERE planet-finder instrument installed at the VLT \citep{sphere2019}, a highly specialised instrument, dedicated to high-contrast imaging and spectroscopy of young giant exoplanets. 
SPHERE is based on the SAXO extreme adaptive optics system \citep{fusco2006, petit2014, sauvage2010}, which controls a deformable mirror with $41\times41$ actuators, and 4 control loops (fast visible tip-tilt, high-orders, near-infrared differential tip-tilt and pupil stabilisation).  The common path optics employs several stress polished toric mirrors \citep{hugot2012} to transport the beam to the coronagraph and scientific instruments. Several types of coronagraphic devices for stellar diffraction suppression are provided, including apodized pupil Lyot coronagraphs \citep{soummer2005} and achromatic four-quadrants phase masks \citep{boccaletti2008}. The instrument has three science subsystems: the infrared dual-band imager and spectrograph (IRDIS, \citealt{dohlen2008}), an integral field spectrograph (IFS; \citealt{claudi2008}) and the Zimpol rapid-switching imaging polarimeter (ZIMPOL; \citealt{zimpol}).

The data were acquired on the 29th January 2018 (Table \ref{tab:obslog}) in IRDIFS-EXT mode, using IRDIS in dual-band imaging (DBI, \citealt{vigan2010}) mode with the $K_1K_2$ filters ($\lambda_{K_1} = 2.1025 \pm 0.1020\,\mu$m; $\lambda_{K_2} = 2.2550 \pm 0.1090\,\mu$m), and IFS in the $Y-H$ ($0.97-1.66\,\mu$m, $R_\lambda=30$) mode in pupil-tracking. This combination enables the use of angular and/or spectral differential imaging techniques to improve the contrast performances at the sub-arcsecond level.

The observing sequence adopted was similar to those designed for the SHINE Guaranteed time survey \citep[see e.g.][]{chauvin2017} and consisted of: 
\begin{itemize}
    \item One PSF sub-sequence composed by a series of off-axis unsaturated images obtained with an offset of $\sim0.4\,''$ relative to the coronagraph center (produced by the Tip-Tilt mirror). A neutral density filter was used to avoid saturation\footnote{\url{www.eso.org/sci/facilities/paranal/instruments/sphere/inst/filters.html}} and the AO visible tip-tilt and high-order loops were closed to obtain a diffraction-limited PSF.
    \item A {\it star center} coronagraphic observation with four symmetric satellite spots, created by introducing a periodic modulation on the deformable mirror \citep[see][for details]{langlois2013}, in order to enable an accurate determination of the star position behind the coronagraphic mask for the following deep coronagraphic sequence. 
    \item The deep coronagraphic sub-sequence, for which we used here the smallest apodized Lyot coronagraph (ALC-YH-S) with a focal-plane mask of 185~mas in diameter.
    \item A new star center sequence, a new PSF registration, as well as a short sky observing sequence for fine correction of the hot pixel variation during the night. 
\end{itemize}

IRDIS and IFS data sets were reduced using the SPHERE Data Reduction and Handling (DRH) automated pipeline \citep{pavlov08} at the SPHERE Data Center \citep[SPHERE-DC, see][]{sphereDC} to correct for each data cube for bad pixels, dark current, flat field and sky background. After combining all data cubes with an adequate calculation of the parallactic angle for each individual frame of the deep coronagraphic sequence, all frames are shifted at the position of the stellar centroid calculated from the initial star center position.
In order to calibrate the IRDIS and IFS data sets on sky, we used images of the astrometric reference field 47 Tuc observed with SPHERE at a date close to our observations. The plate scale and true north values used are reported in Table~\ref{tab:obslog} and are based on the long-term analysis of the GTO astrometric calibration described by \cite{maire2016}. 

The SPHERE-DC corrected products were then processed using the VIP (Vortex Image Processing Package) for High-contrast Direct Imaging \citep{VIP}, which allowed for the speckle pattern subtraction using the angular differential imaging \citep[ADI: ][]{marois2006a} technique within a principal component analysis (PCA) algorithm \citep[see][for details]{soummer2012,amara2012pynpoint}.

\begin{figure}
\includegraphics[width=8cm]{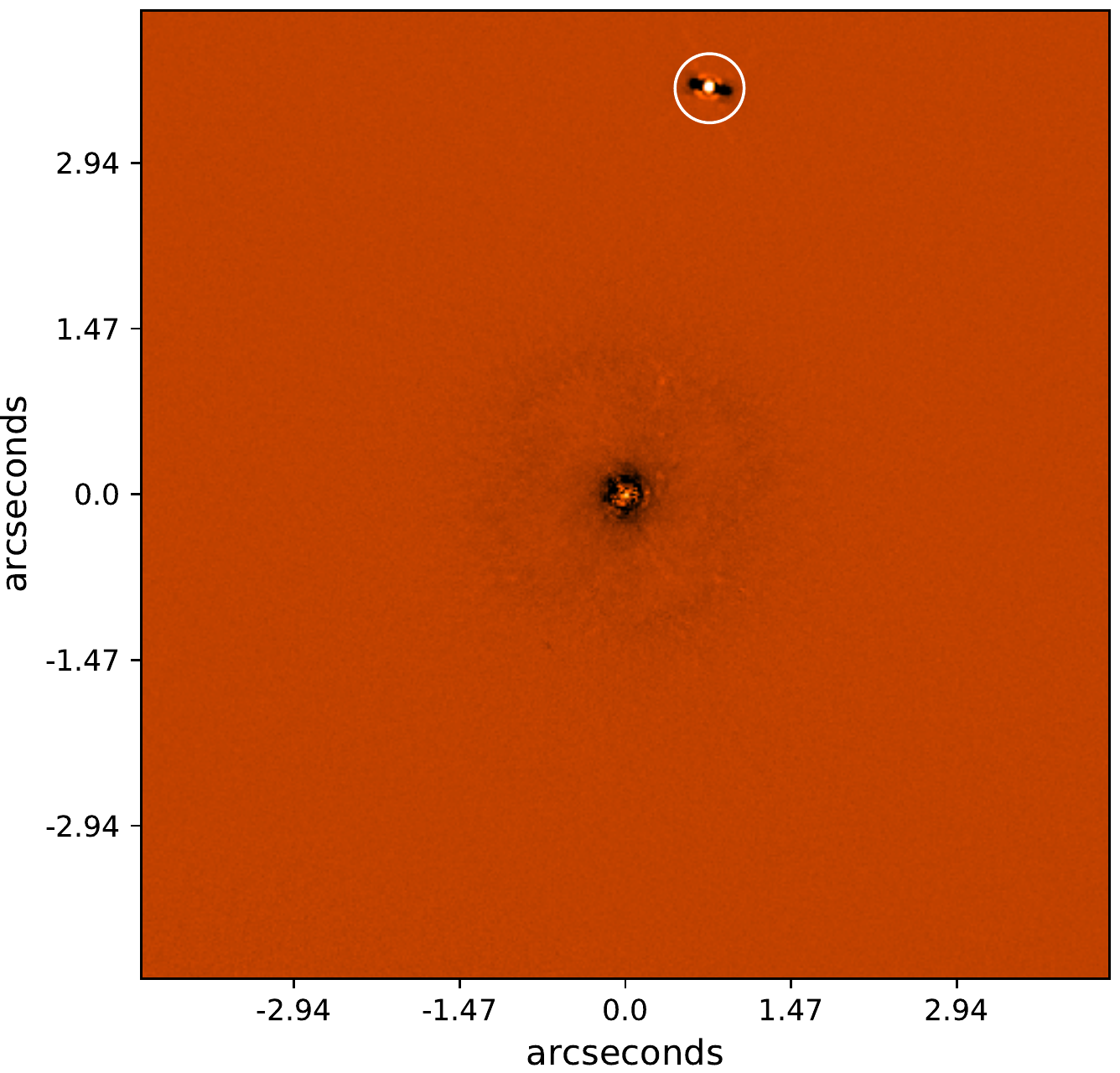}
\caption{IRDIS $K1$-band PCA processed image of GJ 3346 from January 29th 2018. The newly discovered companion GJ 3346~B is highlighted with a white circle.}
\label{fig:fig1}
\end{figure}

\begin{figure}
\begin{centering}
\includegraphics[width=8.2cm]{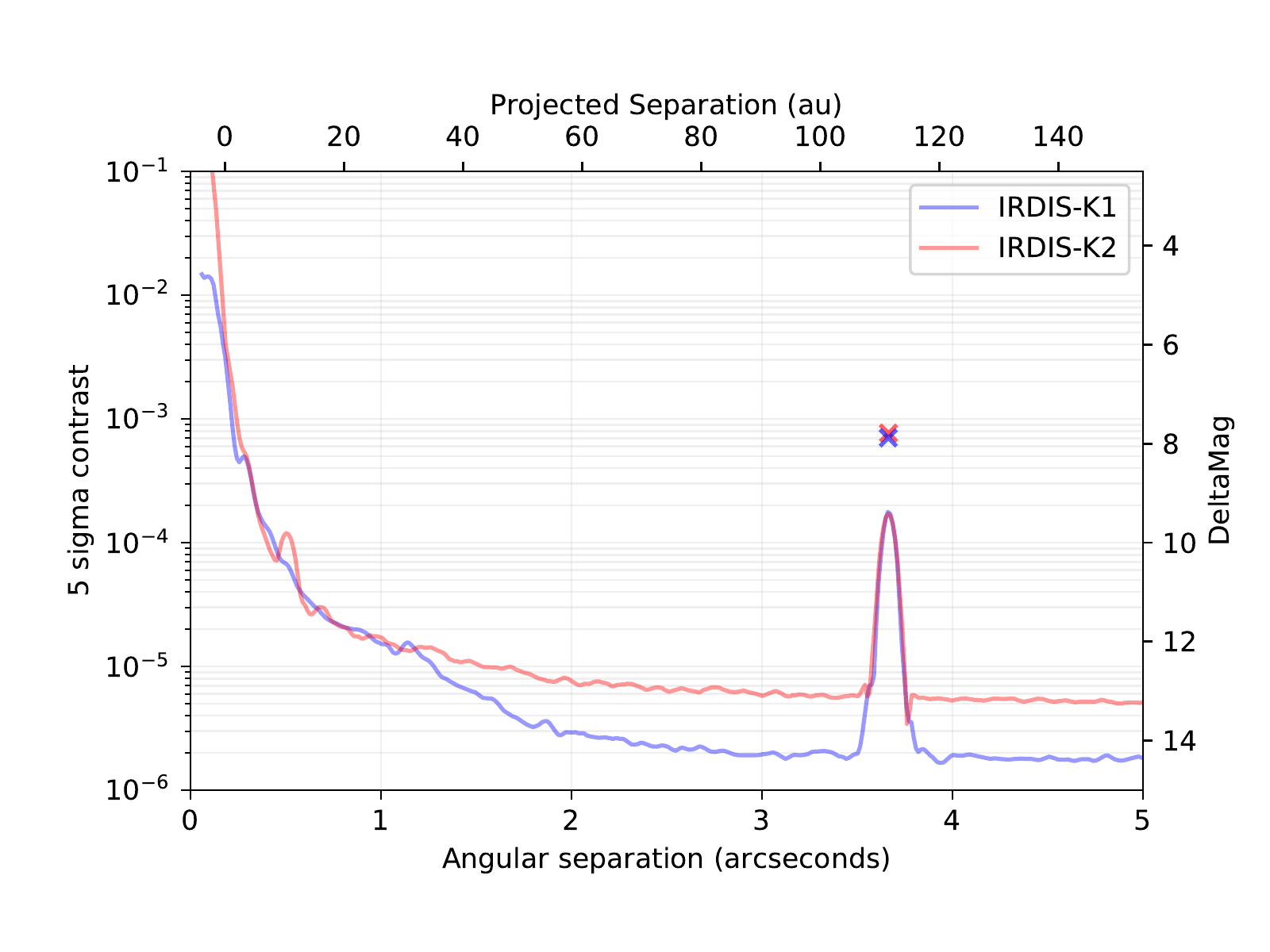}
\caption{5~$\sigma$ contrast limits achieved during the SPHERE observations of GJ 3346 in the two IRDIS filters. The position of GJ~3346~B is marked with a cross.} 
\label{fig:limits}
\end{centering}
\end{figure}

The resulting IRDIS $K_1$ combined image is shown in Fig.~\ref{fig:fig1}, with the position of the newly discovered companion highlighted by the white circle. 
Figure~\ref{fig:limits} shows the corresponding 5$\sigma$ contrast limit as a function of separation from the primary for both IRDIS filters. The bump at $\sim3.6^{\prime\prime}$ is caused by the presence of the companion, which position is marked by the \textit{X} signs. 

\begin{table*}
\caption{Details of VLT/SPHERE observations.}            
\label{tab:obslog}
\begin{tabular}{lllllllllll}     
\hline\hline\noalign{\smallskip}
UT Date    &    Instrument  & Filter   & Pl. Scale &  NDIT$\times$DIT  & $\rm{N}_{\rm{exp}}$      &   Tot.FoV      &  $\omega$    &   Strehl          & Airmass  & TN  \\ 
           &                &          & (mas/pxl) &  (s)              &                       & Rot. ($\degr$)             & ($"$)     &  @1.6$\mu$m    &  &   ($\degr$)       \\
\noalign{\smallskip}\hline\noalign{\smallskip} 
29-01-2018 &    IRDIS       &K1K2      & 12.250  & 1$\times$64       & 17                    & \multirow{2}{*}{17.5}   & \multirow{2}{*}{0.58 }      & \multirow{2}{*}{0.85 }     & \multirow{2}{*}{1.39}   & \multirow{2}{*}{-1.75}  \\
29-01-2018 &    IFS         & Y-H     & 7.46 & 1$\times$64       & 16                    &               &        &       &  &  \\
\noalign{\smallskip} 
\hline\hline\noalign{\smallskip}
\end{tabular}
\end{table*}

\begin{center}
\begin{table}
\caption{Stellar parameters of GJ 3346}\label{tab:schar}
\begin{tabular}{lcl}
\hline\hline
Parameter      & Value  & Reference \\
\hline
Age (Gyr)            &    4.3 - 6.5  & this paper  \\
$M_{star} (M_{\odot})$   &    0.683$\pm$0.018  & this paper \\ 
$R_{star} (R_{\odot})$   &    0.639$\pm$0.014  & this paper \\
\hline 
V (mag)                   &      8.72  & Hipparcos  \\
B$-$V (mag)               &      1.003$\pm$0.003   & Hipparcos  \\
V$-$I (mag)               &      1.05         & Hipparcos \\
G                         &   8.3864 $\pm$ 0.0004 &  GDR2 \\
BP-RP                     &    $0.0640^{+0.0921}_{-0.0561}$ & GDR2 \\
J (mag)                   &     6.856$\pm$0.019  & 2MASS \\
H (mag)                   &     6.284$\pm$0.026  & 2MASS \\
K (mag)                   &     6.205$\pm$0.024  & 2MASS \\
\hline 
RV   (km\,s$^{-1}$)            &  -14.02$\pm$0.42  &  GDR2 \\
U    (km\,s$^{-1}$)            &    -8.24$\pm$0.35  & this paper \\
V    (km\,s$^{-1}$)            &    11.13$\pm$0.27  & this paper \\
W    (km\,s$^{-1}$)            &    30.60$\pm$0.23  & this paper \\
ST                   &   K3V    &  Hipparcos \\
T$_{\mathrm{eff}}$ (K)      &    4750$\pm$65 & this paper \\
$\rm \log g$           &      4.50$\pm$0.10       & this paper \\
$ \rm [Fe/H]    $             &        -0.38$\pm$0.08     & this paper \\
$v \sin i $  (km\,s$^{-1}$)         &   3.5$\pm$0.5 & this paper \\
$P_{rot}$             &   13.0$\pm$0.4             & this paper \\
$\log R_{HK}$         &   -4.48    & \cite{wright2004} \\
$\log L_{X}/L_{bol}$   &  -4.89    &  this paper \\
EW Li (m\AA)         &    0.0   &  this paper \\

\hline 
\multirow{3}{*}{Parallax (mas)} & 41.09 $\pm$ 1.26 & Hipparcos \\
& 42.02 $\pm$ 0.28 & TGAS \\
& 42.0225 $\pm$ 0.0302 & Gaia DR2 \\
\hline 
\multirow{4}{*}{pmRA (mas/yr)}  & 174.3 $\pm$ 1.3 & Tycho-2 \\
& 173.92 $\pm$ 0.87 & Hipparcos \\
& 174.023 $\pm$ 0.064 & TGAS \\
& 173.571 $\pm$ 0.046 & Gaia DR2 \\
\hline
\multirow{4}{*}{pmDEC (mas/yr)} & 201.2 $\pm$ 1.3 & Tycho-2 \\
& 204.52 $\pm$ 0.94 & Hipparcos \\
& 206.393 $\pm$ 0.064 & TGAS \\
& 207.558 $\pm$ 0.053& Gaia DR2 \\

\hline\hline
\end{tabular}
\end{table}
\end{center}

\section{Host star properties } \label{sec:host}

As mentioned in Sec.~\ref{sec:obs}, GJ~3346 was selected as target for our SPHERE program because of significant discrepancies in available measurements of the star's proper motion. 
Table \ref{tab:schar} lists parallax and proper motion measurements found in major astrometric catalogues for GJ~3346. 
A total $\Delta\mu$ of 5.20$\pm$1.84 mas/yr is obtained by comparing the Tycho-2 \citep{tycho} and Tycho Gaia Astrometric Solution (TGAS; \citealt{TGAS}) catalogues, as was done for our selection procedure with COPAINS \citep{fontanive2019b}. A similar value is obtained using the values from the second Gaia Data Release \citep[hereafter GDR2;][]{GDR2} which was not available at the time of target selection, confirming its suitability as a survey target.
In addition to the full five parameter astrometric solution, including celestial position, parallaxes and proper motions, GDR2 also includes photometry in Gaia's G, G$_{BP}$ and G$_{RP}$ bands.
Combining these information with all the available data from the literature, we were able to carefully reassess the values of stellar parameters for GJ~3346, which we report in Table~\ref{tab:schar}, together with the photometry from Gaia, Hipparcos \citep{hip} and 2MASS \citep{2mass}. 

\subsection{Activity and rotation} \label{sec:activity}

\begin{figure}
    \includegraphics[width=7.8cm]{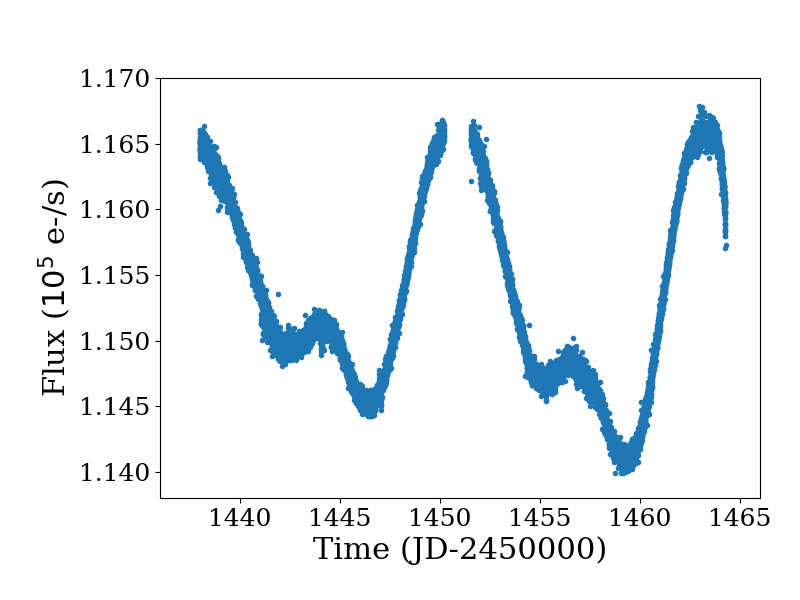}
    \caption{TESS light curve for GJ 3346}
    \label{fig:rotation}
\end{figure}

The star is a K3 star which shows moderate chromospheric activity and X-ray emission.
\cite{wright2004} and \cite{gray2006} measured $\log R_{HK}=-4.48$ and $-4.45$, 
respectively \footnote{The value of $\log R_{HK}$ from \cite{wright2004} is derived from
their tabulated S-Index value using the \cite{noyes1984} prescriptions.}.
An X-ray luminosity of $1.01\times10^{28}$ and $\log L_{X}/L_{bol} = -4.89 $ were derived from ROSAT \citep{rosat_faint} , following the procedures described in \cite{desidera15}.
The availability of TESS \citep{tess} light curves for GJ 3346, shown in Fig.~\ref{fig:rotation}, also allowed us to derive a photometric rotational period of 13.0$\pm$0.4 days\footnote{Data obtained from \url{https://mast.stsci.edu/portal/Mashup/Clients/Mast/Portal.html} - TESS Obs ID: tess2018319095959-s0005-0000000442893646-0125-s}. This is fully compatible with the observed magnetic and coronal activity.

\subsection{Chemical abundances} \label{sec:abundance}
A high-resolution spectrum (spectral coverage from 3800 to 10000 \AA~ with a resolution of R= 57000) of GJ~3346 obtained with FOCES \citep{pfeiffer98} had been published by \cite{maldonado2010}, who claimed a marginal lithium detection, despite the high ($\sim$70) signal-to-noise ratio (SNR) per pixel at the position of the lithium 6708\AA~line.

We used the same data (courtesy of J. Maldonado) to carry out spectroscopic parameter and abundance determination as done in our previous works \citep[see e.g., ][for line lists and solar abundances]{dorazi17}, by using {\sc moog} by C. Sneden (\citeyear{sneden73}, 2017 version) and the ODFNEW grids of model atmospheres (new opacities and no overshooting) by \cite{castelli04}. Atmospheric parameters were derived following the standard procedure: effective temperature (T$_{\mathrm{eff}}$) and micro turbulent velocity ($V_t$) have been obtained by removing spurious trends between $\log$ n(Fe~{\sc i}) and excitation potential and reduced equivalent widths (EWs) of the lines, respectively. 
The surface gravity ($\log g$) comes from the ionisation equilibrium:
\begin{align}
    \Delta [\log(Fe_I~)-\log(Fe_{II})] < \sqrt{\sigma_1^2 + \sigma_2^2}
\end{align}
\noindent where $\sigma_1$ and $\sigma_2$ are errors on the mean abundances from Fe$_{I}$ and Fe$_{II}$, respectively.

We have derived T$_{\mathrm{eff}}=4750\pm$65 K, log$g$=4.5$\pm$0.1 dex, $V_t=0.95 \pm 0.12$ 
km s$^{-1}$, and [Fe/H]=$-0.38 \pm 0.08$ (with the solar iron abundance being $\log n_\odot$(Fe~{\sc i})=7.50). We refer the reader to \cite{dorazi17} for details on error budget computations. 
Our slightly metal-poor iron abundance is consistent with the finding of \cite{mortier2013}
([Fe/H]=-0.20 from CORALIE and \cite{gray2006}, [M/H]=-0.35 from low resolution spectroscopy). Abundances for the $\alpha$ elements Mg, Si, and Ca exhibit a marginal enhancement, suggesting a thin disc composition with [Mg/Fe]=+0.10$\pm$0.12, [Si/Fe]=+0.07$\pm$0.09, and [Ca/Fe]=+0.19$\pm$0.11 dex.

As a possible indication of pollution from the previous AGB companion we have also derived the s-process element Ba, to search for enhancements. However, we did not detect any hint of over-abundance with [Ba/Fe]=+0.05 $\pm$ 0.10 (see \cite{dorazi17} and references therein for details on Ba abundance determination).
The quality of the FOCES spectrum does not allow us investigate in detail the occurrence of significant alterations of abundances other key elements, such as e.g., carbon, yttrium, zirconium, and lanthanum. However, the solar Ba abundance suggests that this is not the case and further investigations, by acquiring new high quality (SNR $\gtrsim$ 150), high resolution spectra are not crucial in this context.  

\begin{figure}
    \includegraphics[width=7.8cm]{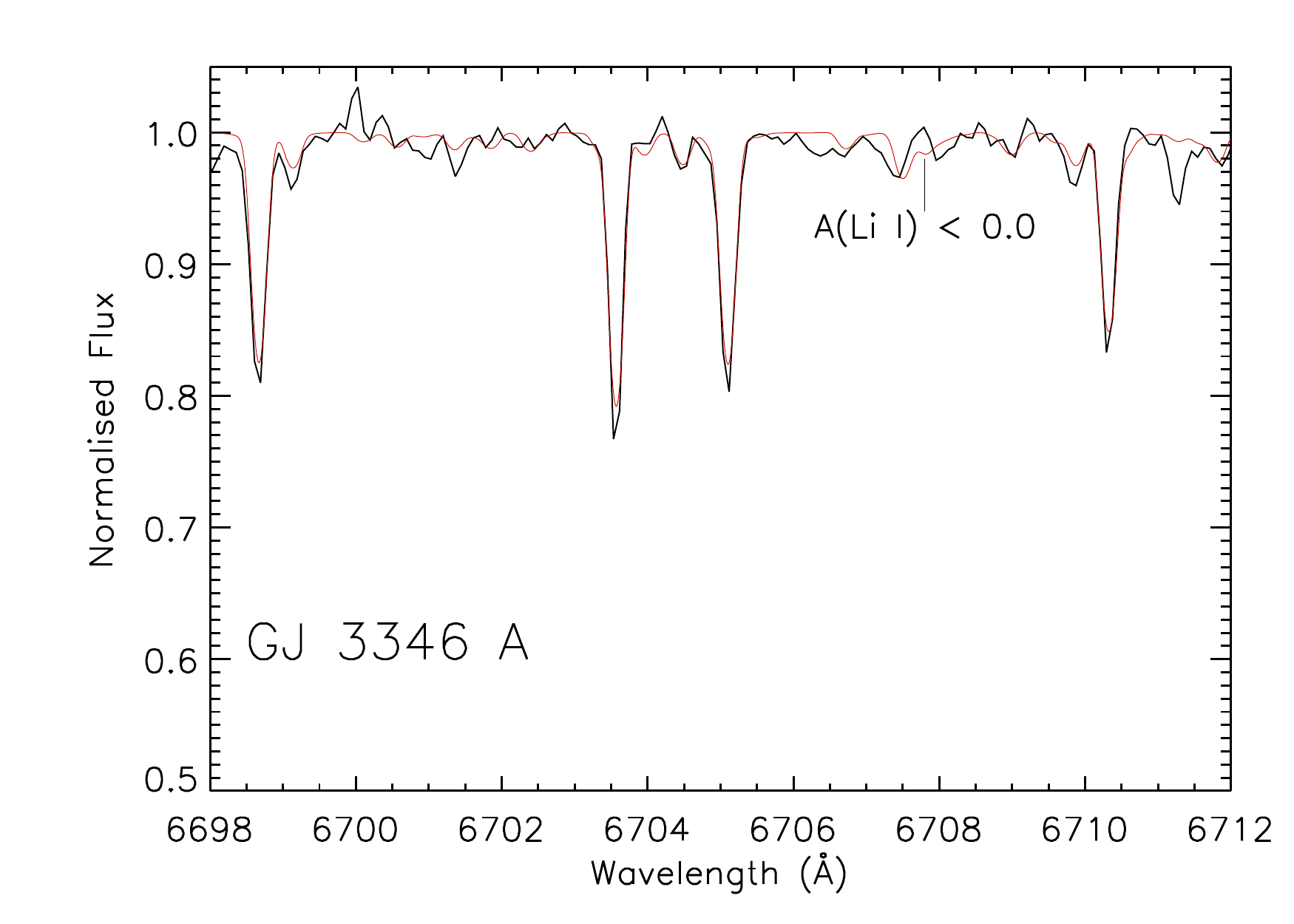}
    \caption{Spectral synthesis (red line) around the Li~{\sc i} line at 6707.8 \AA~ for GJ 3346 A.}
    \label{fig:lithium}
\end{figure}

As shownd in Fig.~\ref{fig:lithium}, a comparison between the observed and synthetic spectrum (calculated assuming A(Li)=0.00) only allows to put an upper limit on the lithium abundances. We believe that the measurement reported by \cite{maldonado2010} is likely a blended equivalent width of the iron line.
From the spectral synthesis including the BaII (5853 \AA) and Li spectral regions we also derived the projected rotational velocity of the star (v$\sin i$=3.5$\pm$0.5 km/s). 

\subsection{System age}\label{sec:age}
GJ~3364 was originally classified as young based on the information on the rotation and chromospherical and coronal emission.
Our revised values of these indicators, discussed in Sec.~\ref{sec:activity}, are still compatible with an age of $\sim$ 600-700 Myr for GJ~3346. They could, however, also be explained assuming it is instead an older star rejuvenated by angular momentum accreted through stellar wind originated from the WD progenitor at the end of the AGB phase \citep[see][and references therein]{zurlo2013} or by tidal locking with a close stellar companion \citep[see e.g.][]{hut1981, fleming2019}.

The possibility of a tidally locked binary can easily be ruled out by the availability in the literature of several RV measurements \citep{nordstrom2004,maldonado2010,sperauskas2016,riedel2017,Gaia_Main}.
These data show some scatter exceeding the formal errors (peak-to-valley differences of about 3  km/s) but the dispersion is low enough to exclude a tidally-locked binary as source for the moderate magnetic activity of the star.

The lack of a clear lithium detection, on the other hand, also points towards an older age for the system, with the WD being responsible for the spin up. 
This is not surprising considering that the projected separation of the WD (87 au) is similar to the one observed for HD8049 (50 au). HD 8049 is a system with a K-type central star, a WD companion and significant signature of rejuvenation \citep{zurlo2013}.
In the case of GJ 3346, the activity is significantly lower than that of HD8049.
We expect this is due to an earlier event of mass loss from the WD progenitor and subsequent accretion. If this was the case, we would have expected a significantly longer cooling age for the WD around GJ~3346.

The relatively low metallicity discussed in Sec.~\ref{sec:abundance} is another indication for an old age since nearby young stars typically have chemical composition close to solar \citep{dorazi11, biazzo17}, which is also supported by the stellar kinematics. 
While kinematic arguments do not provide a well defined age (a part from the case of members of moving groups or associations with well defined age), it can still be used to obtain robust limits.
Using Gaia DR2 astrometric values and absolute RV  we obtain U,V,W = -8.24, 11.12, 30.60 km/s. 
The W velocity is well outside the kinematic space of young stars \citep{montes2001} and U and V velocities are also marginally inconsistent with it. This confirms that the star is older than 1 Gyr.

We finally tried to derive the age of the system through isochrone fitting, using the PARAM\footnote{\url{http://stev.oapd.inaf.it/cgi-bin/param_1.3}} \citep{param} web interface and adopting spectroscopic T$_{\mathrm{eff}}$ and [Fe/H], GDR2 parallax and V band magnitude. 
The resulting value of  5.3$\pm$3.4 Gyr is inconclusive, as expected for a K dwarf close to the main sequence. The stellar mass resulting from this fit is of 0.683$\pm$0.018 $M_{\odot}$.

Given the ambiguities described above, we decided to adopt a different approach and tried to assess the most probable system age by estimating the typical values for stars with kinematics and metallicity similar to those of GJ 3346.
We selected from \cite{casagrande2011} the stars with metallicity and galactic orbit similar to GJ~3346 (eccentricity between 0.04 to 0.10; maximum height over the galactic plane between 0.70 to 0.82, [Fe/H] between -0.2 to -0.6). All the objects with blended photometry were then removed, which yielded a sample of 13 objects (beside GJ~3346), none of which detected in X-ray or with reported signatures pointing towards an age lower than 1 Gyr in the literature.
We finally derived the stellar ages for these targets using PARAM, adopting effective temperature and [Fe/H] from \citealt{casagrande2011}), obtaining a median value of the age is 5.3 Gyr with a dispersion of 2.5 Gyr. 
The median values of minimum and maximum ages obtained using the error bars provided by the PARAM web interface are 4.3 and 6.5 Gyr respectively. We chose to use these as the adopted age range of the system, as this is still consistent with the estimated isochrone age for GJ3346, but more accurate.

\section{Results}\label{sec:res}

\subsection{Detection of a co-moving WD companion to GJ~3346}
A point source at a separation of 3.665$\pm$0.002 arcsec and a position angle of 348.31$\pm$0.09 deg from GJ~3346 was retrieved in our IRDIS images (see Fig.~\ref{fig:fig1}). 
A source compatible with the candidate identified in the IRDIS field of view was also retrieved in Gaia DR2 ( $\rho$=3.647$\pm$0.001 arcsec, PA=347.89$\pm$0.02 deg, see Tab.~\ref{tab:photoastro} for details). 
The fact that the GDR2 parallax and proper motion of this object were very similar to those of GJ~3346 provided a strong indication of its co-moving nature, despite the apparent discrepancies within the single values. 
In fact, as discussed in \cite{fontanive2019a}, differences in both parallax and proper motion such as the ones observed here are to be expected in kinematics measurements made over a short time span, as is the case for GDR2 parameters, which capture the reflex orbital motions in the components of multiple systems. Indeed, for $\Delta\mu$ binaries, short-term proper motions will by definition be deviant from the center-of-mass motion of the pair, and in different directions for the two components at opposite ends of their orbits.
In addition, the presence of the bright primary within few arcseconds almost certainly affected the quality of the Gaia DR2 5-parameter astrometric solution for GJ~3346~B, which appearsto be relatively poor (its Renormalised Unit Weight Error\footnote{RUWE, described in details in \url{http://www.rssd.esa.int/doc_fetch.php?id=3757412}} is in fact $\sim$4, as opposed to the typical value of 1.4 expected for a good fit.)

To further confirm the co-moving nature of GJ~3346~B, we estimated the expected motion of a background star relative to GJ~3346 over the 2.5 yrs baseline between our SPHERE observation and Gaia DR2, given the parallax and proper motion of the primary. 
The results, plotted in Fig.~\ref{fig:cpm}, clearly show that the measured positions (reported in Tab.~\ref{tab:photoastro}) are incompatible with a background source, thus validating the idea that the pair is physically associated.

\begin{table*}
\caption{Relative photometry and astrometry of GJ 3346 B}             
\label{tab:photoastro}
\centering
\begin{tabular}{lllccc}     
\hline\hline\noalign{\smallskip}
Epoch & Filter  & $\lambda_c$     &  Contrast       & Separation         &   PA                                         \\
      &         & ($\mu m$)       & ($\Delta$mag)   & (mas)              &    ($\degr$)                      \\
\hline
\multirow{3}{*}{2015.5}  &  Gaia G  & 639.74  & 5.829 $\pm$ 0.005 & \multirow{3}{*}{3647.55 $\pm$ 1.00} & \multirow{3}{*}{347.8935 $\pm$ 0.02} \\
                         & Gaia BP  & 516.47  & 4.228 $\pm$ 0.150   &  &  \\
                         & Gaia RP  & 783.05  & 4.686 $\pm$ 0.136   &  &  \\
\multirow{2}{*}{2018.05} & IRDIS K1 & 2.1025  & 7.844 $\pm$ 1.628 & 3665.59 $\pm$ 1.77 & 348.30 $\pm$ 0.07      \\
                         & IRDIS K2 & 2.2550  & 7.788 $\pm$ 0.057 & 3664.45 $\pm$ 1.42 & 348.33 $\pm$ 0.06      \\
\hline\hline\noalign{\smallskip}
\end{tabular}
\end{table*}

\subsection{Companion photometric characterisation}

\begin{figure}
\includegraphics[width=9.5cm]{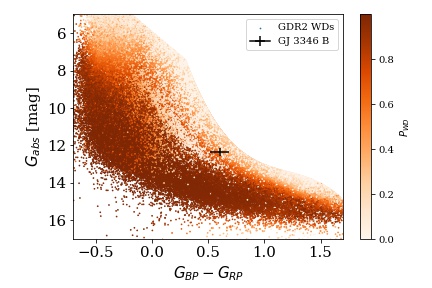}
\caption{Position of GJ 3346~B (black square) in the GDR2 WD locus, as per the catalogue compiled by \citet{gentile2019}. The colour scale indicate the probability of the source being a WD, $P_{WD}$, as defined in the catalogue description.}
\label{fig:gdr2_wds}
\end{figure}

\begin{figure}
\includegraphics[width=0.45\textwidth]{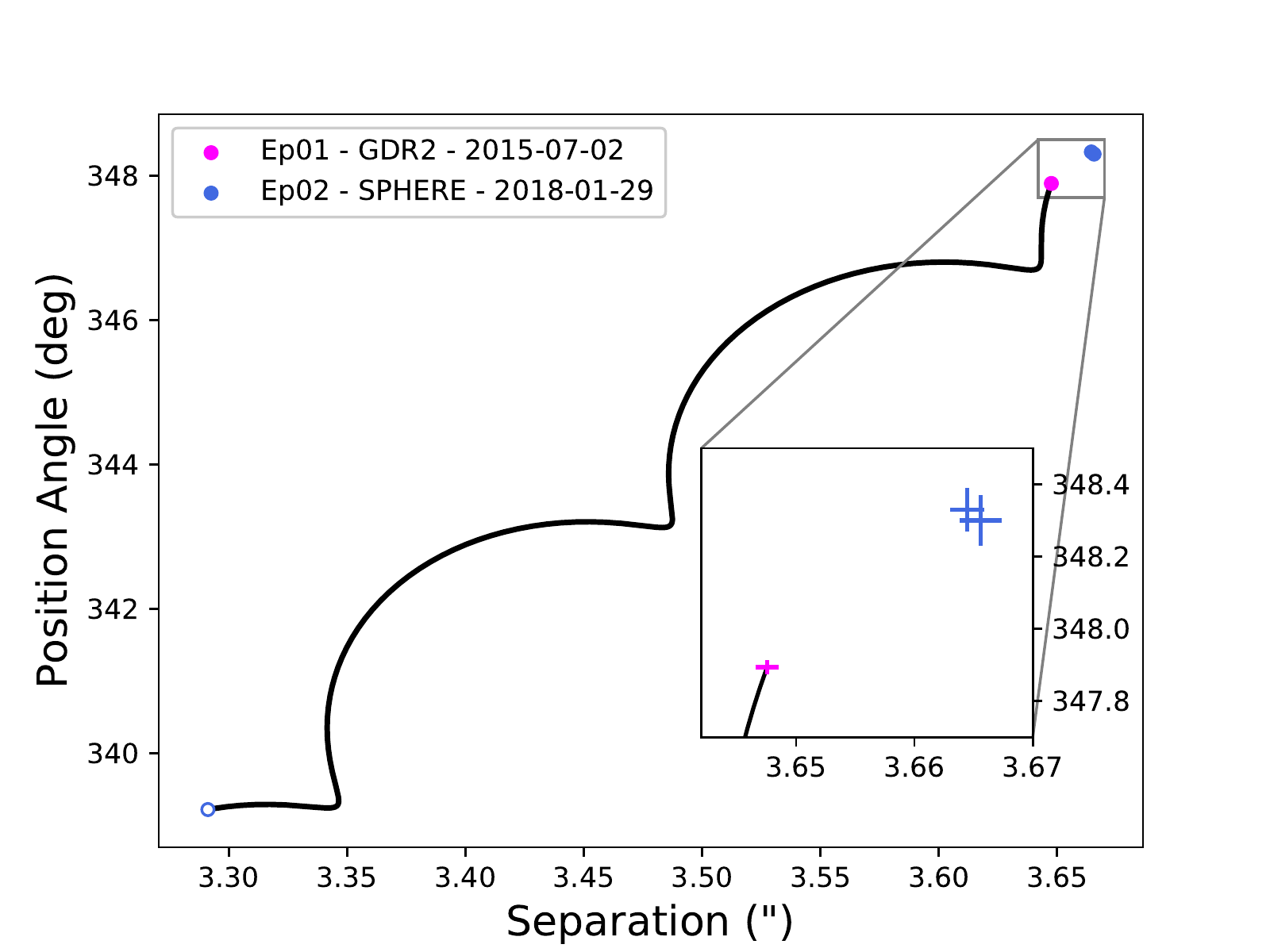}
\caption{Common proper motion analysis of GJ 3346 and its companion over the $\sim$2.5 year baseline between GDR2 (magenta) and the astrometry from our SPHERE data (blue). The black line shows the motion of a background object relative to GJ 3346 based on the GDR2 parallax and proper motion of the primary over the same time frame, and the blue open circle indicates the expected position of a background object at the epoch of the SPHERE detection. The close companion is clearly found to be co-moving with our target.}
\label{fig:cpm}
\end{figure}

Table~\ref{tab:photoastro} shows the values of the photometry of GJ 3346 B from IRDIS using the VIP package \citep{VIP} and those retrieved from the GAIA DR2 catalogue.
While the red colour in the IRDIS bands pointed towards a sub-stellar nature for the companion, the photometry from Gaia lead us to think GJ 3346~B could be a WD instead. 
When we compared its colours with the ones of the objects in the WD locus from \cite{GaiaWD} (see Fig.~\ref{fig:gdr2_wds}), 
we found it to be compatible with the WD sequence considering the large colour error\footnote{The large 
$phot\_bp\_rp\_excess\_factor$ (3.783) indicates a significant contamination
by the primary on the BP-RP colour \citep{evans2018}. 
The true colour is then likely bluer, better placing GJ 3346 B  on the WD sequence.}.
The blue colour of the companion is further supported by comparison of Gaia and SPHERE photometry, that yields G-K1 and G-K2 equal to 0.1266 and 0.2226 respectively. 
This is a further confirmation of  the WD nature of GJ~3346~B.
There is no available photometry in the UV from Galex, while U band photometry of the whole system shows no or small UV excess, depending on the adopted U band photometry \citep{koen2010,mermilliod1997}, thus excluding a very hot source.

\subsection{Derived physical parameters}
\label{sec:wd} 

As in \cite{zurlo2013}, we used a catalogue of empirical sequences using the catalogue of nearby WD{s} by \cite{2012ApJS..199...29G}. The photometry values have been supplemented with available GALEX $FUV$ and $NUV$ magnitudes, and 2MASS $J$, $H$ and $K_S$ magnitudes. The final sample consists of 107 nearby ($\leq 51$~pc) WDs: 22 with $FUV$ magnitude{s}, 18 with $NUV$ magnitude{s} and 84 with $J$, $H$, $K_S$ magnitudes calibrated by \cite{2012ApJS..199...29G}. Along with the empirical catalogue we used the theoretical sequences of \cite{2011MNRAS.410.2095V}.

\begin{figure*}
\centering
\includegraphics[width=8.82cm]{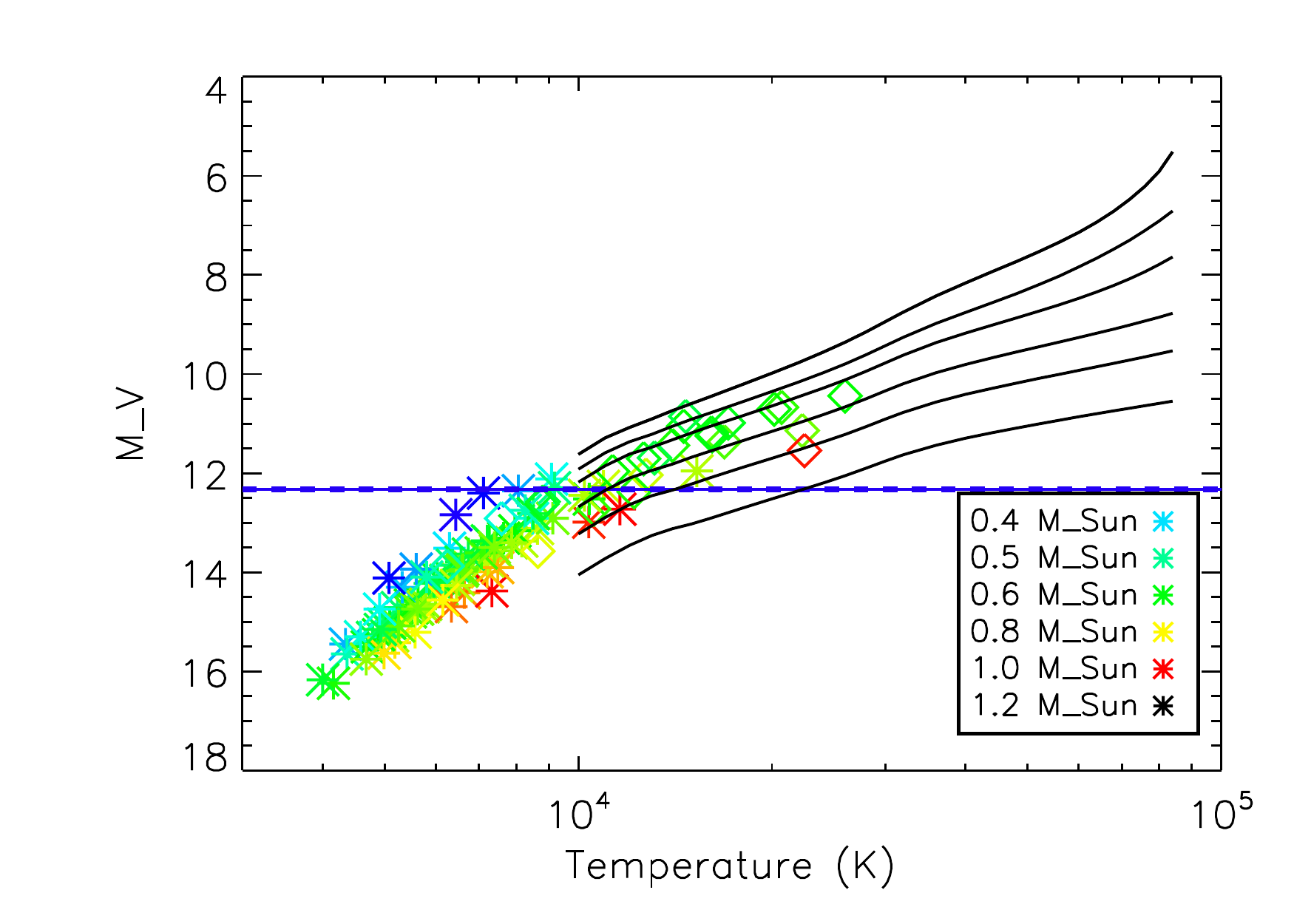}
\includegraphics[width=8.82cm]{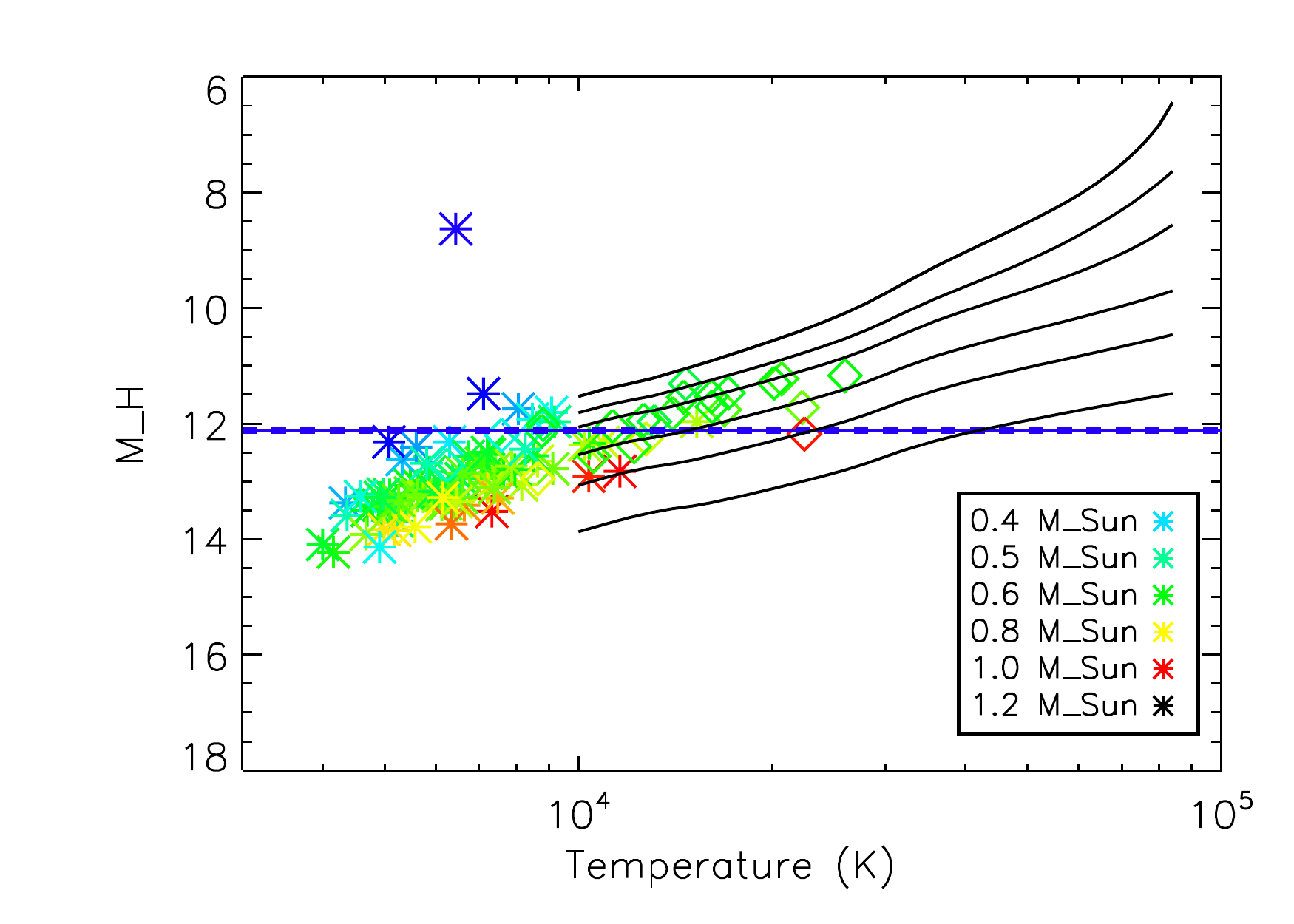}

\includegraphics[width=8.82cm]{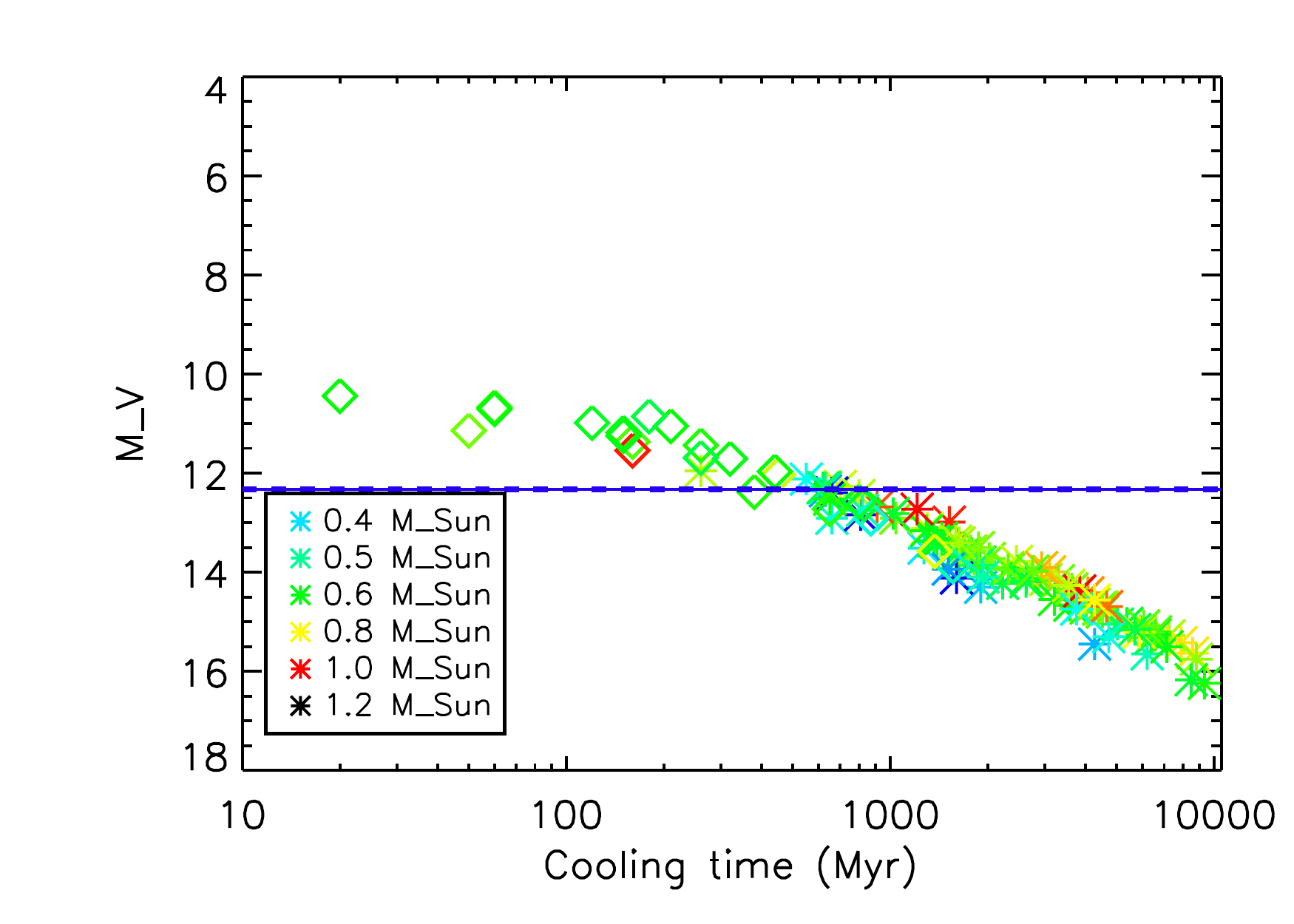}
\includegraphics[width=8.82cm]{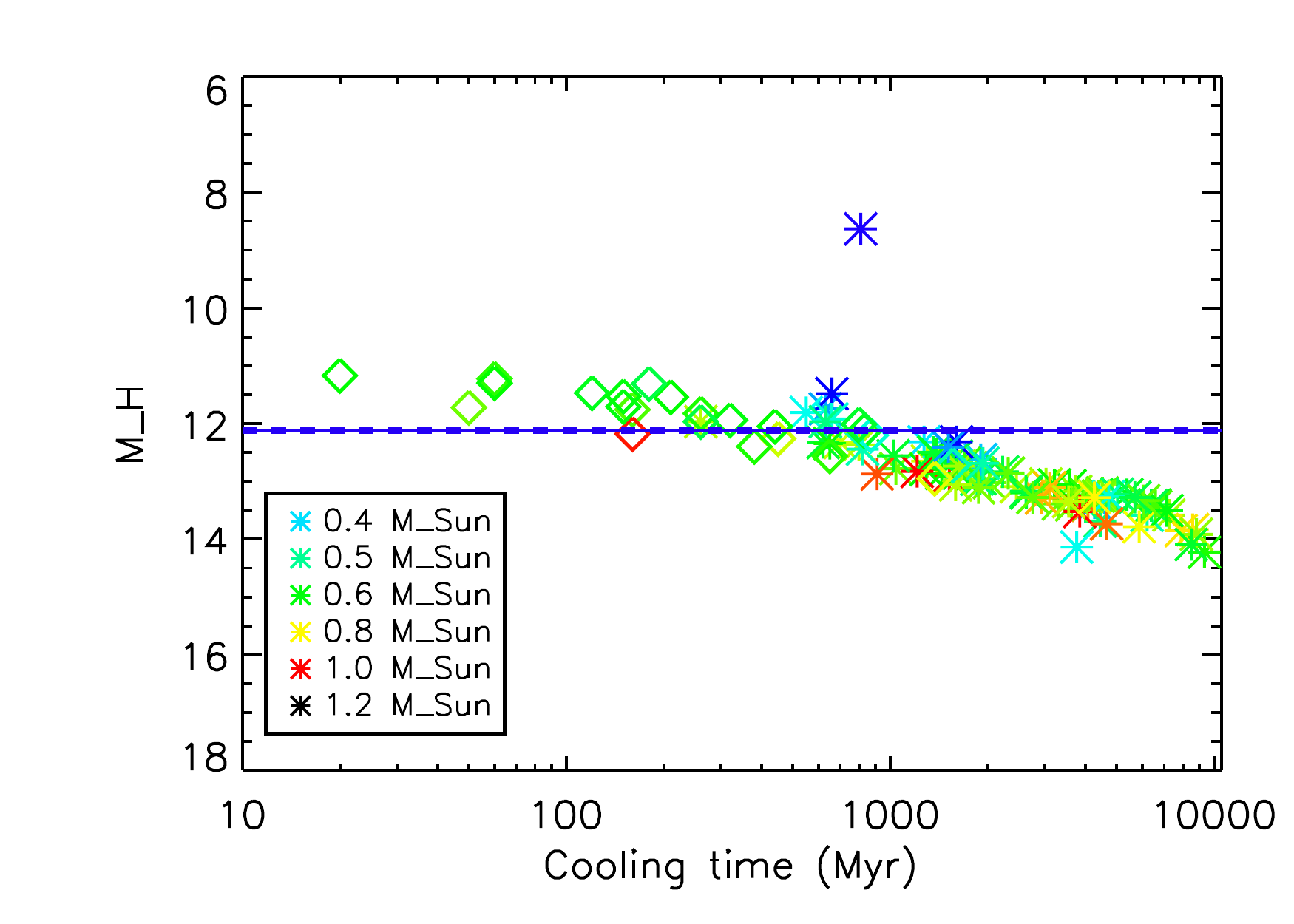}
\caption{{{\it Top. }Absolute magnitude in different bands ($V$ and $H$) versus effective temperature for the WD models of \citet{2011MNRAS.410.2095V} (black lines) and a sample of nearby dwarfs collected by \citet{2012ApJS..199...29G}.  Stars represent objects with all magnitudes available and diamonds represent the others. Colours indicate the mass from the lowest (blue) to the highest (red ones). The continuous blue horizontal line represents the magnitude of GJ~3346~B. The corresponding error bars are plotted as dashed blue lines. The plots show that the objects of the \citet{2012ApJS..199...29G} catalogue are not peculiar and are well described by the theoretical models of \citet{2011MNRAS.410.2095V}.} {\it Bottom.} Absolute $V$ and $H$ magnitude versus cooling age of the WD. }
\label{f:wdage}
\end{figure*}

We calculated the effective temperature T$_{\mathrm{eff}}$ using the empirical and theoretical sequences from the visible and NIR photometry of the WD. The result is a temperature of $\sim 11 \times 10^{3}$ K (see Fig.~\ref{f:wdage}, top). This imply a a value of the mass of the WD between 0.45 and 0.7 $M_{\odot}$. 
The cooling time calculated using the empirical sequences is $684 \pm 58$~Myr. 
(see Fig.~\ref{f:wdage}, bottom).

\subsection{System history}
\label{sec:history} 


Coupling the estimates of the total system age and of the WD cooling time allows us some inferences of the most probable original configuration and evolution of the system.
To this aim, we used the pre-WD lifetimes from the \cite{bressan2012} models for the appropriate metallicity and the initial-final WD mass relationship by \cite{cummings2018}.
Subtracting the WD cooling age from the system age yield a most probable pre-WD lifetime of 4.6 Gyr, with plausible limits between 3.6 to 5.8 Gyr. The corresponding initial masses are 1.20, 1.30, and 1.12 $M_{\odot}$, respectively,
corresponding to WD masses of 0.585$\pm$0.008 $M_{\odot}$ for the adopted relationship.

Neglecting the small amount of material accreted by the K-type component after the mass loss and assuming adiabatic expansion of the orbit \citep{huang1956,boffin2015}, one could expect the original separation to have been roughly two thirds of the present one, i.e. 58 au. The true semi major axis could be different due to orbit eccentricity and on-sky projection effects. 

Younger/older system ages, corresponding to more/less massive WD progenitors, would imply a tighter/wider original configuration, respectively.
While speculative, this evolution of the system would explain quite naturally all the observed features, including the rejuvenation of the K type component through wind accretion \citep{jeffries1996}.

\begin{center}
\begin{table}
\caption{Summary of the properties of GJ 3346~B}\label{tab:binchar}
\begin{tabular}{lcl}
\hline\hline
Parameter      & Value  & Reference \\
\hline
Projected separation (au) & 86.50$\pm$0.06 & this work \\
Current Mass ($M_{\odot}$) & 0.58 $\pm$ 0.01 & this work \\
\hline 
IRDIS-K1 mag    & 14.04$\pm$ 1.64 & this work \\
IRDIS-K2 mag    & 13.99$\pm$ 0.07 & this work \\
Gaia G mag      & 14.22$\pm$0.01 & GDR2 \\ 
Gaia BP mag     & 13.19$\pm$0.15 & GDR2 \\ 	
Gaia RP mag     & 12.59$\pm$0.14 & GDR2 \\ 
\hline 
Parallax (mas)  & $42.30\pm0.07$  & GDR2\\  
pmRA (mas/yr)   & $182.24\pm0.10$ & GDR2 \\  
pmDEC (mas/yr)  & $216.25\pm0.12$ & GDR2 \\  
\hline 
T$_{\mathrm{eff}}$ (K) &$11\times10^3 \pm $ 500 & this work \\
Cooling time (Myr) & 684$\pm$58  & this work \\
Main sequence time (Gyr) & $4.6^{+1.2}_{-1.0}$  & this work \\
Original Mass ($M_{\odot}$) & $1.20^{+0.10}_{-0.08}$ & this work \\
\hline\hline
\end{tabular}
\end{table}
\end{center}

\subsection{Astrometric trend due to the WD companion}
\label{sec:astrometry}

\begin{figure*}
\includegraphics[width=8.0cm]{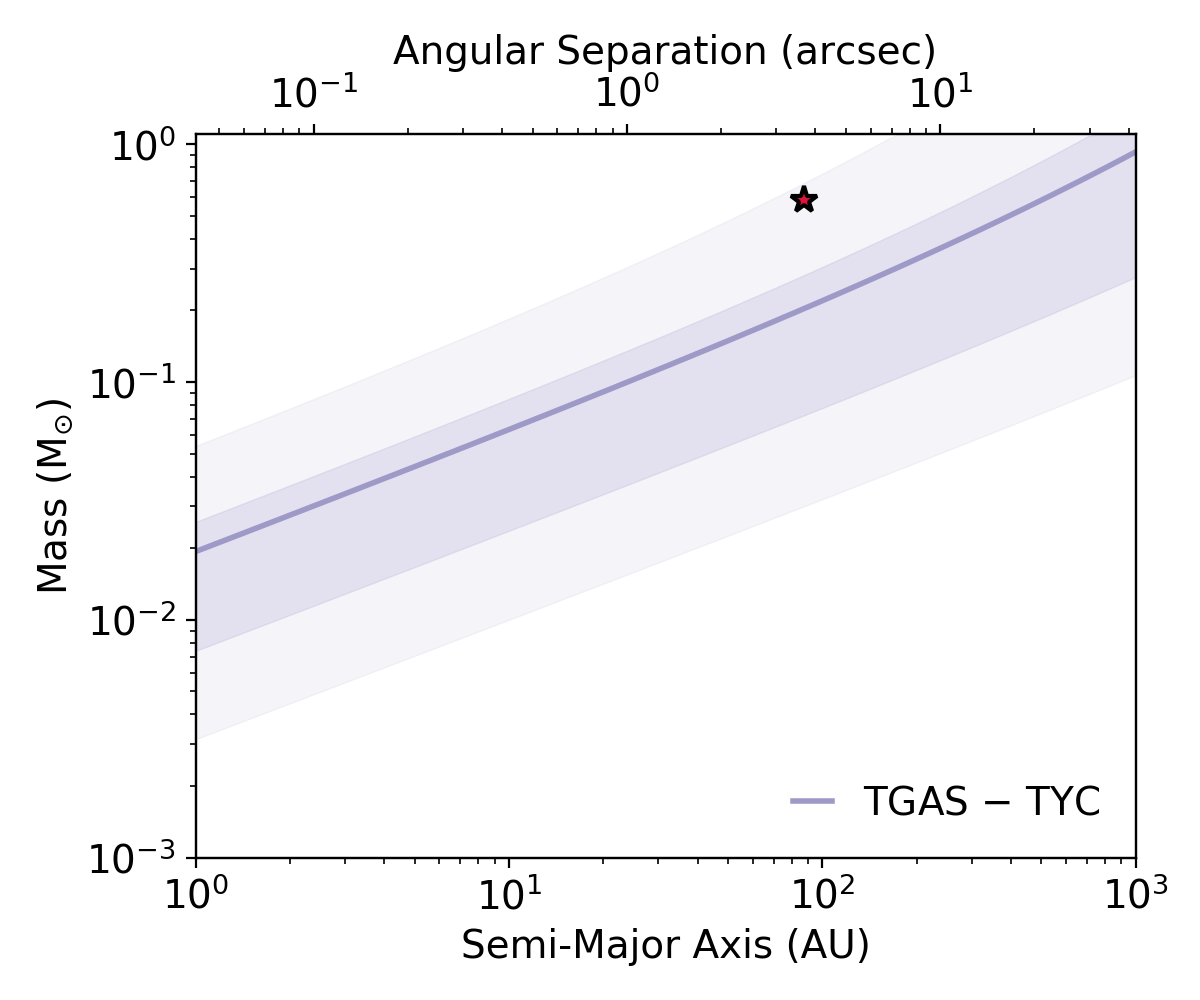}
\includegraphics[width=8.0cm]{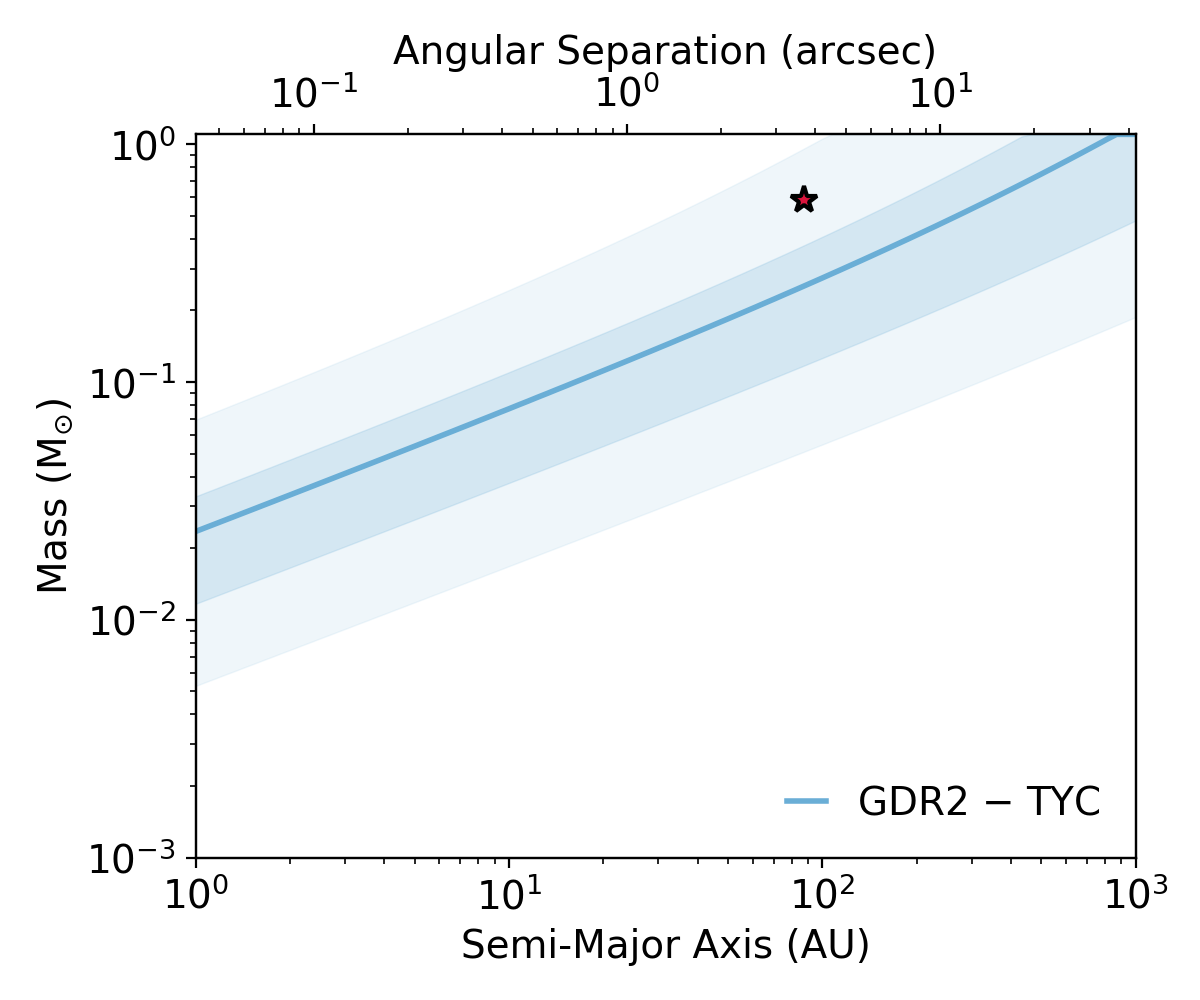}
\caption{Output of the COPAINS code for GJ 3346, showing predictions of the possible solutions for the mass and separation of the companion inducing the observed $\Delta\mu$ trend of the primary. The left panel shows the expected position of the companion using a combination of the TGAS and Tycho-2 proper motions of the star, as used in the original target selection process, showing the median (solid line) and 1 and 2-$\sigma$ intervals (dark and light shaded regions) of the possible solutions. The right panel shows the same predictions made subsequently with GDR2 astrometry. The position of the detected WD companion is marked by the red star, compatible at the 1.5$-$2 $\sigma$ level with the expectations. We note however that the separation of the companion corresponds to a projected separation, while the dynamical predictions are in semi-major axis.}
\label{fig:deltamu_plots}
\end{figure*}

As previously noted, GJ 3346 has a strong $\Delta\mu$ offset between Tycho-2 and TGAS proper motions (Table \ref{tab:schar}). Assuming that the Tycho-2 measurement is close to the centre-of-mass motion of the system, and that Gaia provides a good approximation to the instantaneous velocity of the star, the COPAINS tool enables predictions of the possible masses and separations of the secondary companion. The analysis conducted with COPAINS on GJ 3346 revealed that the companion triggering the astrometric trend could be compatible with a substellar secondary on separations smaller than a few tens of AU, or with a more massive stellar companion on a larger orbital distance. The results from the predictions made with COPAINS are shown in Fig.~\ref{fig:deltamu_plots}. The solid line shows the median set of solutions for the position of the bound companion, and the dark and light envelopes represent the 1 and 2$\sigma$ regions of confidence, respectively. The predictions assume a flat distribution in eccentricity and a face-on orbit.

As TGAS has a $\sim$25-yr baseline, this catalogue will only be a good estimate of short-term proper motions for systems with orbital periods on the century timescale. As a result, the predictions made from TGAS measurements may not be accurate at small orbital separations (left panel). Nonetheless, the identified companion being on a wide orbit, this is not expected to affect our system. The position of the WD, indicated by the red star in Fig. \ref{fig:deltamu_plots}, is indeed in agreement with the expectations from COPAINS within 2$\sigma$.

After the release of the Gaia DR2 catalogue, the same analysis was repeated with GDR2 for completeness and is shown in the right panel. With a baseline of 22 months only, GDR2 truly captures the reflex motion of the star under the gravitational influence of the WD, and provides an excellent approximation to an instantaneous velocity. The fact that the dynamical predictions in the two panels of Fig. \ref{fig:deltamu_plots} are very similar confirms that the timescale of TGAS was also very short relative to the orbital period of the GJ 3346 AB system. The more accurate and better trusted COPAINS simulations made with GDR2 proper motions are consistent with the measured mass and observed separation of GJ 3346 B at the 1.5$\sigma$ level.

\section{Discussion and conclusions}
\label{s:conclusion}

We detected a new WD companion around the K-type star GJ 3346, observed with SPHERE as part
of the COPAINS pilot survey (Bonavita et al. in prep) a program focused on young stars with significant proper motion difference between Gaia DR1 and Tycho-2.
The companion was first detected in the SPHERE observations and then retrieved in Gaia DR2, which allowed to both confirm its comoving nature and to identify it as a WD companion. 

Compared to similar systems discovered with similar methods \citep[such as HD~8049, see][for details]{zurlo2013}, GJ~3346 lacked the abundance of information from the literature (in particular, there are no UV data), resulting in a less accurate estimate of the parameters of the WD and the possible original configuration of the system.
We were nonetheless able to constrain its effective temperature ($11 \times 10^3 \pm 500$ K) and cooling age (684$\pm$58 Myr).

The analysis of the central K star shows that the young age originally  inferred from rotation and chromospheric and coronal emission is refuted by the lack of lithium, moderately low metallicity, and kinematics. The magnetic activity can be explained by spin-up of the star thanks to accretion of material and angular momentum at the end of the AGB phase of the WD component.
Interestingly, the age from rotation and activity is close to the estimated cooling age of the WD, as found  by \cite{zurlo2013} for HD 8049 and by \cite{leiner2018} for some other systems.
This further supports the idea that accretion of angular momentum by the progenitor of the WD at the of the AGB phase reset the rotational clock of the star and then the rotational evolution proceeded as isolated stars after the accretion.

While the detection of the companion confirms the validity of our selection method, GJ3346~B is not suitable for a detailed analysis to obtain an estimate of its dynamical mass, as done for other objects showing significant $\Delta\mu$, such as HD~284149~B \citep{bonavita2017}. Given its long period and the lack of information on its inclination and eccentricity\footnote{Given that the available RV were obtained with different instruments, is not possible to obtain a reliable combination and to further constrain orbital parameters and minimum mass. Furhter observations at a precision of about 10 m/s spanning few years would be required to uncover a RV trend.}, such analysis, based on the method proposed by \cite{makarov2005} could only lead to a rather unreliable estimate of its minimum mass at best (Fontanive et al., in prep., for a detailed description of the mass estimate method as well as its limitations).

With a projected separation of 87 au, GJ~3346~B is expected to cause an RV trend of $\sim$ 15-30~ms$^{-1}$yr$^{-1}$ \citep[estimated following the approach by][]{liu2002}, which is well within reach of high-precision RV instruments, making it a promising case for detection. Further astrometric constraints on its mass will most likely also be provided by future Gaia releases.
This discovery therefore confirms the key role played by Gaia in the discovery and characterisation of faint companions, as well as the potential of dynamical pre-selection methods to enhance the yield of direct imaging surveys.

\section*{Acknowledgements}
Based on observations made with ESO Telescopes at the La Silla Paranal Observatory under programme ID 100.C-646A.\\
We thank the anonymous referee for extensive feedback that significantly improved the clarity of the paper.\\
We thank J. Maldonado for kindly us providing the reduced FOCES spectrum.\\
M.B. and B.B acknowledge funding by the UK Science and Technology Facilities Council (STFC) grant no. ST/M001229/1.\\
M.B., C.F. and K.M. acknowledge funding by the Carnegie Trust Research Incentive Grant no. RIG007779.\\
S.D., V.D., D.M. and R.G. acknowledge the support by INAF/Frontiera through the "Progetti Premiali" funding scheme of the Italian Ministry of Education, University, and Research. \\
K.M. acknowledges funding by the Science and Technology Foundation of Portugal (FCT), grants No. IF/00194/2015 and PTDC/FIS-AST/28731/2017. \\
A.Z. acknowledges support from the CONICYT + PAI/ Convocatoria nacional subvenci\'on a la instalaci\'on en la academia, convocatoria 2017 + Folio PAI77170087. \\
This work has made use of the the SPHERE Data Centre, jointly operated by OSUG/IPAG (Grenoble), PYTHEAS/LAM/CESAM (Marseille), OCA/Lagrange (Nice), Observatoire de Paris/LESIA (Paris), and Observatoire de Lyon. \\
Some of the data presented in this paper were obtained from the Mikulski Archive for Space Telescopes (MAST). STScI is operated by the Association of Universities for Research in Astronomy, Inc., under NASA contract NAS5-26555. Support for MAST for non-HST data is provided by the NASA Office of Space Science via grant NNX13AC07G and by other grants and contracts.




\bibliographystyle{mnras}
\bibliography{bib}




\bsp	
\label{lastpage}
\end{document}